# Error Exponents and Cutoff Rate for Noncoherent Rician Fading Channels

Mustafa Cenk Gursoy
Department of Electrical Engineering
University of Nebraska-Lincoln, Lincoln, NE 68588
Email: mgursoy@unl.edu

*Abstract*— In this paper, random coding error exponents and cutoff rate are studied for noncoherent Rician fading channels, where neither the receiver nor the transmitter has channel side information. First, it is assumed that the input is subject only to an average power constraint. In this case, a lower bound to the random coding error exponent is considered and the optimal input achieving this lower bound is shown to have a discrete amplitude and uniform phase. If the input is subject to both average and peak power constraints, it is proven that the optimal input achieving the random coding error exponent has again a discrete nature. Finally, the cutoff rate is analyzed, and the optimality of the single-mass input amplitude distribution in the low-power regime is discussed.

## I. INTRODUCTION

There has recently been much progress in the design of wireless systems, striving for high data rates achieved at low levels of probability of error. These advances has motivated the information-theoretic study of fading channels, which provides not only the ultimate performance limits but also the design guidelines to achieve these limits. One information-theoretic notion is the channel capacity. The capacity of fading channels has been extensively studied (see e.g., [2]). However, if neither the transmitter nor the receiver has channel side information, the channel capacity is in general not known except for several special cases (see e.g., [5], [7]) where the capacity does not have a closed-form expression and has to be computed using numerical methods. For noncoherent fading channels, general results are available only in the asymptotically high-SNR [3] and low-SNR [4] regimes.

In this paper, we study other information-theoretic notions, namely error exponents and cutoff rate. Our main focus will be on random coding error exponents. In [1], Gallager derived upper bounds on the probability of error that can be achieved by block codes on general discrete-time memoryless channels. Using an ensemble of codebooks where each letter of each codeword is chosen independently of all other letters with a certain probability distribution, it is shown in [1] that for any rate $R$ less than the channel capacity, the probability of error can be upper bounded by

$$P_e \leq B \exp(-NE(R)) \qquad (1)$$

where $B$ is a constant, $N$ is the codeword length, and $E(R)$ is the random coding error exponent. $E(R)$ provides the interactions between the probability of error, channel coding, data rates, and the signal-to-noise ratio (SNR). The cutoff rate $R_0$ provides a lower bound to the channel capacity. Although turbo codes are shown to achieve rates higher than the cutoff rate, for sequential decoding strategies, $R_0$ is the maximum practical symbol rate [8]. Moreover in cases where channel capacity is not known, cutoff rate may prove to be useful providing achievable rates.

Recently, there has been interest in obtaining the error exponents and cutoff rate for noncoherent channels where neither the receiver nor the transmitter knows the fading coefficients. Hero and Marzetta [8] characterized the structure of input signals achieving the cutoff rate in unknown Rayleigh fading multiple antenna channels subject to peak power constraints, and investigated its implications on the signal design. Lapidoth and Miliou [10] studied the error exponents and obtained the high SNR expansion of the cutoff rate in Rician channels. Abou-Faycal and Hochwald [9] analyzed the random coding error exponents for unknown Rayleigh fading channels, and shown that the optimal input achieving the random coding error exponent in single-antenna Rayleigh block-fading channels has a discrete character. More recently, Huang *et al.* [11] proved the discreteness of the optimal input achieving the random coding error exponent for a general class of channels when there is only a peak power constraint.

## II. CHANNEL MODEL

We consider the following Rician fading channel model

$$y_i = dx_i + a_i x_i + n_i \quad i = 1, 2, \ldots \qquad (2)$$

where $x_i$ is the complex channel input, $y_i$ is the channel output, $d$ is a complex constant, $\{a_i\}$ and $\{n_i\}$ are sequences of independent and identically distributed (i.i.d.) zero mean circularly symmetric complex Gaussian random variables, representing the fading coefficients and additive background noise samples respectively. It is further assumed that $a_i$ and $n_i$ are independent, and $E\{|a_i|^2\} = \gamma^2$ and $E\{|n_i|^2\} = N_0$. Therefore, the conditional output probability density function given the input is

$$f_{y|x}(y|x) = \frac{1}{\pi(\gamma^2|x|^2 + N_0)} \exp\left(-\frac{|y - dx|^2}{\gamma^2|x|^2 + N_0}\right). \qquad (3)$$

In (2), fading has a multiplicative effect. Hence, it is assumed that the delay spread of the channel is much less than the symbol duration. Moreover, since the fading coefficients are assumed to be independent for each sample, the channel is memoryless. Under these fast fading conditions, the noncoherent scenario, where neither the receiver nor the transmitter knows the fading coefficients, is considered.

The Rician fading is a suitable model for wireless channels where there exits, in addition to the multipath fading components, a line-of-sight link between the receiver and transmitter. Furthermore, the Rician fading channel serves as a unifying model including both the Rayleigh fading and the unfaded Gaussian channels as special cases.

## III. OPTIMAL INPUT STRUCTURE FOR THE RANDOM CODING ERROR EXPONENTS

### A. Average Power Limited Input

In this section, we study the structure of the optimal input achieving the random coding error exponent when the input is subject to an average power constraint,

$$E\{|x_i|^2\} \leq P \quad \forall i. \tag{4}$$

The random coding error exponent is obtained from the following optimization problem

$$E(R) = \sup_{0 \leq \rho \leq 1} \sup_{\nu \geq 0} \sup_{\substack{F_x \\ E\{|x|^2\} \leq P}} \left( E_0(\rho, F_x, \nu) - \rho R \right) \tag{5}$$

where

$$E_0(\rho, F_x, \nu) = -\log \int_y \left( \int_x e^{\nu(|x|^2 - P)} f_{y|x}(y|x)^{\frac{1}{1+\rho}} \, dF_x(x) \right)^{1+\rho} dy. \tag{6}$$

Due to the monotonicity of $\log(\cdot)$, the optimal choice of the input distribution function and the parameter $\nu$ can equivalently be obtained by solving

$$L(\rho) = \inf_{\nu \geq 0} \inf_{\substack{F_x \\ E\{|x|^2\} \leq P}} L_0(\rho, F_x, \nu) \tag{7}$$

where

$$L_0(\rho, F_x, \nu) \triangleq \int_y \left( \int_x e^{\nu(|x|^2 - P)} f_{y|x}(y|x)^{\frac{1}{1+\rho}} \, dF_x(x) \right)^{1+\rho} dy. \tag{8}$$

First, we have the following Lemma which characterizes the optimal input phase distribution.

*Lemma 1:* An input with uniformly distributed phase that is independent of the amplitude achieves $L(\rho)$ in (7).

*Proof*: Assume that that an input $x$ achieves $L_0(\rho, F_x, \nu)$. Consider a new input $\hat{x} = x e^{j\theta}$ where $\theta$ is independent of $x$ and uniformly distributed over $[-\pi, \pi)$. Assume further that $\hat{x}$ achieves $L_0(\rho, F_{\hat{x}}, \nu)$. Using the property that $f_{y|x}(ye^{j\theta}|xe^{j\theta}) = f_{y|x}(y|x)$ for fixed $\theta$, it can easily be seen that $L_0(\rho, F_{\hat{x}}, \nu|\theta) = L_0(\rho, F_x, \nu)$ where $L_0(\rho, F_{\hat{x}}, \nu|\theta)$ denotes the functional value conditioned on $\theta$. Since $L_0$ is a convex function of the input distribution for $\rho \in [0, 1]$, we can show using Jensen's inequality that $L_0(\rho, F_{\hat{x}}, \nu) \leq L_0(\rho, F_x, \nu)$. Note also that if $x$ satisfies the power constraint, then so does $\hat{x}$ which has uniformly distributed phase that is independent of the amplitude. Therefore, we do not lose optimality by assuming uniformly distributed phase, and hence the Lemma follows. □

With the above characterization, the optimization problem in (7) has been reduced to finding the optimal input amplitude distribution and the optimal $\nu$. We define the following random variables:

$$r = \frac{\gamma |x|}{\sqrt{N_0}} \quad \text{and} \quad R = \frac{|y|^2}{N_0}. \tag{9}$$

Using these definitions and integrating with respect to the uniform input phase, $L_0(\rho, F_x, \nu)$ becomes

$$L_0(\rho, F_x, \nu) = \int_0^\infty \left( \int_0^\infty g(R, r) \, dF_r(r) \right)^{1+\rho} dR \tag{10}$$

$$\triangleq L_0(\rho, F_r, \bar{\nu}) \tag{11}$$

where

$$g(R, r) = \frac{\exp(\bar{\nu}(r^2 - \alpha))}{(1 + r^2)^{\frac{1}{1+\rho}}} \exp\left( -\frac{R + \mathsf{K}r^2}{(1+\rho)(1+r^2)} \right) \times I_0 \left( \frac{2\sqrt{\mathsf{K}} r \sqrt{R}}{(1+\rho)(1+r^2)} \right) \tag{12}$$

with $\bar{\nu} = \frac{\nu N_0}{\gamma^2}$, $\alpha = \frac{\gamma^2 P}{N_0}$, and $\mathsf{K} = \frac{|d|^2}{\gamma^2}$ which is the Rician factor. $I_0$ is the zeroth-order modified Bessel function of the first kind. Now, the optimization problem in (7) can be rewritten as

$$L(\rho) = \inf_{\bar{\nu} \geq 0} \inf_{\substack{F_r \\ E\{r^2\} \leq \alpha}} \int_0^\infty \left( \int_0^\infty g(R, r) \, dF_r(r) \right)^{1+\rho} dR \tag{13}$$

We now choose the suboptimal value $\bar{\nu} = 0$[1] and concentrate on the problem

$$\hat{L}(\rho) \triangleq \inf_{\substack{F_r \\ E\{r^2\} \leq \alpha}} T(F_r) \tag{14}$$

$$= \inf_{\substack{F_r \\ E\{r^2\} \leq \alpha}} \int_0^\infty \left( \int_0^\infty \hat{g}(R, r) \, dF_r(r) \right)^{1+\rho} dR \tag{15}$$

where

$$\hat{g}(R, r) = \frac{1}{(1 + r^2)^{\frac{1}{1+\rho}}} \exp\left( -\frac{R + \mathsf{K}r^2}{(1+\rho)(1+r^2)} \right) \times I_0 \left( \frac{2\sqrt{\mathsf{K}} r \sqrt{R}}{(1+\rho)(1+r^2)} \right) \tag{16}$$

Since the optimization problem in (14) is over continuous alphabets, the existence of an optimal input achieving the infimum is not immediate. Therefore we first note the following result.

*Theorem 1:* For all $\rho \in [0, 1]$, there exists an input amplitude distribution that achieves the infimum in (14).

*Proof*: The existence of an optimal distribution is proved if the input distribution function space over which the minimization is performed is compact, and the objective functional is weak* continuous [13]. The compactness of the space of input distributions with second moment constraints is shown in [5]. Therefore, we need only to show the weak* continuity of $T(\cdot)$. The weak* continuity of the functional $T(\cdot)$ is equivalent to

$$F_n \xrightarrow{w^*} F \Rightarrow T(F_n) \to T(F). \quad (17)$$

---

[1]The discreteness result is also obtained considering $\bar{\nu} = 0$ in [11].

Since $\hat{g}(R, r)$ is a continuous and bounded function for all $r \geq 0$ and $R \geq 0$, by the definition of weak convergence [13],

$$F_n \xrightarrow{w^*} F \Rightarrow \int_0^\infty \hat{g}(R, r) \, dF_n(r) \to \int_0^\infty \hat{g}(R, r) \, dF(r) \quad (18)$$

for all $R \geq 0$. Then,

$$\lim_{n \to \infty} \int_0^\infty \left( \int_0^\infty \hat{g}(R, r) \, dF_n(r) \right)^{1+\rho} dR \quad (19)$$

$$= \int_0^\infty \lim_{n \to \infty} \left( \int_0^\infty \hat{g}(R, r) \, dF_n(r) \right)^{1+\rho} dR \quad (20)$$

$$= \int_0^\infty \left( \lim_{n \to \infty} \int_0^\infty \hat{g}(R, r) \, dF_n(r) \right)^{1+\rho} dR \quad (21)$$

$$= \int_0^\infty \left( \int_0^\infty \hat{g}(R, r) \, dF(r) \right)^{1+\rho} dR \quad (22)$$

In the above formulation, (21) follows from the continuity of the function $x^{1+\rho}$, and (22) follows from (18). Hence, if interchange of the limit and integration is justified in (20), the weak* continuity of $T(\cdot)$ is proved. To justify (20), we can invoke the Dominated Convergence Theorem which requires an integrable upper bound on the integrand

$$h(R, F_n) = \left( \int_0^\infty \hat{g}(R, r) \, dF_n(r) \right)^{1+\rho}.$$

Following an analysis similar to that in [6], we can show that

$$h(R; F_n) \leq \begin{cases} 1 & 0 \leq R < 2 \\ \min\left\{1, \left(e^{\frac{-R^{3/4} + \sqrt{K}R^{1/2}}{1+\rho}} + \frac{a\alpha}{R^{\frac{1}{1+\rho}}(R^{1/4}-1)}\right)^{1+\rho}\right\} & R \geq 2. \end{cases} \quad (23)$$

Note that the right-hand side of (23) does not depend on $n$ and is integrable because it decreases as $O\left(\frac{1}{R^{1+\frac{1+\rho}{4}}}\right)$ for large values of $R$. □

Next, we show a sufficient and necessary condition for an input distribution to achieve the minimum in (14).

*Theorem 2:* (*Kuhn-Tucker condition*) An input distribution $F_0$ achieves the minimum in (14) if and only if there exists $\lambda \geq 0$ such that

$$\Phi(r) = \int_0^\infty \left( \int_0^\infty \hat{g}(R, r) \, dF_r(r) \right)^\rho \hat{g}(R, r) \, dR$$

$$- T(F_0) + \lambda \frac{r^2 - \alpha}{1+\rho} \geq 0 \quad \forall r \geq 0 \quad (24)$$

with equality if $r \in E_0$ where $E_0$ is the set of points of increase of $F_0$[2].

The proof of the sufficient and necessary condition in Theorem 2 follows along the same lines as those shown in [5]

[2]The set of points of increase of a distribution function $F$ is $\{r : F(r-\epsilon) < F(r+\epsilon) \; \forall \epsilon > 0\}$.

and [7]. The main steps of the proof is comprised of forming the Lagrangian and then evaluating the weak derivative defined as

$$T'_{F_0}(F) \triangleq \lim_{\theta \to 0} \frac{T[(1-\theta)F_0 + \theta F] - T(F_0)}{\theta}. \quad (25)$$

Note that the if $F_0$ is indeed the minimizing distribution, then $T'_{F_0}(F) \geq 0$ for all $F$ satisfying the power constraint. The Kuhn-Tucker condition will be employed to show that the optimal input amplitude distribution has a discrete nature.

*Theorem 3:* Fix $\rho \in [0, 1]$. If the average power constraint (4) is an active constraint, then the optimal input amplitude distribution that achieves the minimum in (14) is discrete with a finite number of mass points.

*Proof*: Assume $F_0$ is an optimal input satisfying the Kuhn-Tucker condition (24). In order to prove the discreteness of $F_0$, we first obtain a lower bound on the left-hand side of (24). To that end, we initially have

$$\int_0^\infty \hat{g}(R, r) \, dF_r(r) \geq D_{F_0} e^{-\frac{R}{1+\rho}} \quad (26)$$

where

$$D_{F_0} = \int_0^\infty \frac{1}{(1+r^2)^{\frac{1}{1+\rho}}} e^{-\frac{Kr^2}{(1+\rho)(1+r^2)}} \, dF_0(r) > 0. \quad (27)$$

The lower bound in (26) follows easily by noting that

$$e^{-\frac{R}{(1+\rho)(1+r^2)}} \geq e^{-\frac{R}{(1+\rho)}} \quad \text{and} \quad I_0(r) \geq 1 \quad \forall r \geq 0. \quad (28)$$

This bound immediately leads to

$$\int_0^\infty \left( \int_0^\infty \hat{g}(R, r) \, dF_r(r) \right)^\rho \hat{g}(R, r) \, dR \quad (29)$$

$$\geq \int_0^\infty D_{F_0}^\rho e^{-\frac{R\rho}{1+\rho}} \hat{g}(R, r) \, dR \quad (30)$$

$$\geq \frac{D_{F_0}^\rho}{(1+r^2)^{\frac{1}{1+\rho}}} e^{-\frac{Kr^2}{(1+\rho)(1+r^2)}} \int_0^\infty e^{-R} \, dR \quad (31)$$

$$\geq \frac{D_{F_0}^\rho}{(1+r^2)^{\frac{1}{1+\rho}}} e^{-\frac{Kr^2}{(1+\rho)(1+r^2)}}. \quad (32)$$

Note that (31) follows from

$$\hat{g}(R, r) \geq \frac{1}{(1+r^2)^{\frac{1}{1+\rho}}} e^{-\frac{Kr^2}{(1+\rho)(1+r^2)}} e^{-\frac{R}{1+\rho}} \quad \forall R, r \geq 0. \quad (33)$$

Using (32), we obtain the following lower bound on the left-hand side (LHS) of (24):

$$\Phi(r) \geq \frac{D_{F_0}^\rho}{(1+r^2)^{\frac{1}{1+\rho}}} e^{-\frac{Kr^2}{(1+\rho)(1+r^2)}} - T(F_0) + \lambda \frac{r^2 - \alpha}{1+\rho}. \quad (34)$$

It is readily observed that the above lower bound diverges to infinity as $r \uparrow \infty$ when $\lambda > 0$, i.e., the average power constraint is active. Employing the techniques used in [7], we can show that if the optimal input amplitude distribution has an infinite number of points of increase on a bounded interval, then the Kuhn-Tucker condition (24) should be satisfied with equality for all $r \geq 0$. Note that this condition is satisfied by continuous distributions. If the input is discrete with an infinite number of mass points but only finitely many of them on any bounded interval, the Kuhn-Tucker condition should be satisfied infinitely often as $r \to \infty$. Clearly, neither of these conditions can be fulfilled because of the diverging lower bound in (34). Hence, the optimal input must be discrete with a finite number of mass points. $\square$

*B. Average and Peak Power Limited Input*

In this section, we assume that the input is subject to both average and peak power constraints

$$E\{|x_i|^2\} \leq P \quad \text{and} \quad |x_i|^2 \stackrel{\text{a.s.}}{\leq} \kappa P \quad \forall i \tag{35}$$

where $\kappa \geq 1$. Note that $\kappa$ can be seen as a limitation imposed on the peak-to-average power ratio. In this case, the random error exponent is given by

$$E(R) = \sup_{0 \leq \rho \leq 1} \sup_{\nu \geq 0} \sup_{\substack{F_x \\ E\{|x|^2\} \leq P \\ |x|^2 \leq \kappa P}} \left( E_0(\rho, F_x, \nu) - \rho R \right) \tag{36}$$

where $E_0(\rho, F_x, \nu)$ is given by (6). Similarly as in the previous section, uniform phase is optimal, and the optimization problem can be recast as

$$L(\rho) = \inf_{\bar{\nu} \geq 0} \inf_{\substack{F_r \\ E\{r^2\} \leq \alpha \\ r^2 \leq \kappa \alpha}} \int_0^\infty \left( \int_0^\infty g(R,r) \, dF_r(r) \right)^{1+\rho} dR \tag{37}$$

where $g(R,r)$ is given by (12). Note that $g(\cdot, \cdot)$ is a continuous and bounded function for all $r \in [0, \sqrt{\kappa \alpha}]$ and $R \geq 0$. Hence, the existence of the optimal input amplitude distribution achieving the infimum in (37) can be easily shown for any $0 \leq \rho \leq 1$ and $\bar{\nu} \geq 0$, following the approach employed in Section III-A. Indeed, the proof is more straightforward as $g(R,r)$ is exponentially decreasing as a function of $R$ in the presence of a peak power constraint. For this case, the sufficient and necessary (Kuhn-Tucker) condition is given by

$$\Phi_p(r) = \int_0^\infty \left( \int_0^\infty g(R,r) \, dF_r(r) \right)^\rho g(R,r) \, dR$$
$$- T(F_0) \geq 0 \tag{38}$$

for all $r \in [0, \sqrt{\kappa \alpha}]$ with equality if $r \in E_0$ where $E_0$ is the set of points of increase of $F_0$. Here,

$$T(F_0) = \int_0^\infty \left( \int_0^\infty g(R,r) \, dF_0(r) \right)^{1+\rho} dR. \tag{39}$$

It should be emphasized that (38) is a sufficient and necessary condition for both the optimal input amplitude distribution $F_0$ and the optimal value of $\bar{\nu}$. In fact, (38) is obtained by incorporating

$$(1+\rho) \int_0^\infty \left( \int_0^\infty g(R,r) \, dF_r(r) \right)^\rho \int_0^\infty (r^2 - \alpha) g(R,r) dF(r) dR = 0 \tag{40}$$

where the LHS is the derivative of $T$ with respect to $\bar{\nu}$, into (24) to yield $\lambda = 0$. The discreteness of the optimal amplitude can similarly be proved using the following lower bound

$$\int_0^\infty \left( \int_0^\infty g(R,r) \, dF_r(r) \right)^\rho g(R,r) \, dR \tag{41}$$

$$\geq \int_0^\infty D_{F_0}^\rho e^{-\frac{R\rho}{1+\rho}} g(R,r) \, dR \tag{42}$$

$$\geq \frac{D_{F_0}^\rho}{(1+r^2)^{\frac{1}{1+\rho}}} e^{-\frac{\mathsf{K}r^2}{(1+\rho)(1+r^2)}} e^{\bar{\nu}(r^2-\alpha)} \int_0^\infty e^{-R} \, dR \tag{43}$$

$$\geq \frac{D_{F_0}^\rho}{(1+r^2)^{\frac{1}{1+\rho}}} e^{-\frac{\mathsf{K}r^2}{(1+\rho)(1+r^2)}} e^{\bar{\nu}(r^2-\alpha)}. \tag{44}$$

where

$$D_{F_0} = \int_0^\infty \frac{e^{\bar{\nu}(r^2-\alpha)}}{(1+r^2)^{\frac{1}{1+\rho}}} e^{-\frac{\mathsf{K}r^2}{(1+\rho)(1+r^2)}} \, dF_0(r) > 0. \tag{45}$$

Note that the lower bound in (44) diverges as $r \uparrow \infty$ when $\bar{\nu} > 0$. It can be shown similarly as in Section III-A that if the optimal input has an infinite number of points of increase on a bounded interval, $\Phi_p(r) = 0$ for all $r \geq 0$ which is not possible due to the above lower bound. Hence the optimal input amplitude distribution must be discrete. Note that for the discreteness result, we again require the average power constraint to be active, i.e., $\bar{\nu} > 0$.

IV. CUTOFF RATE ANALYSIS IN THE LOW-POWER REGIME

When the input is subject only to an average power constraint, the cutoff rate is given by

$$R_0 = \sup_{\nu \geq 0} \sup_{\substack{F_x \\ E\{|x|^2\} \leq P}} E_0(\rho, F_x, \nu)|_{\rho=1} \tag{46}$$

$$= \sup_{\nu \geq 0} \sup_{\substack{F_x \\ E\{|x|^2\} \leq P}} -\log \int_y \left( \int_x e^{\nu(|x|^2 - P)} \sqrt{f_{y|x}(y|x)} \, dF_x(x) \right)^2 dy. \tag{47}$$

As in the previous section, we choose the suboptimal value $\nu = 0$, and consider

$$\hat{R}_0 = \sup_{\substack{F_x \\ E\{|x|^2\} \leq P}} -\log \int_y \left( \int_x \sqrt{f_{y|x}(y|x)} \, dF_x(x) \right)^2 dy. \tag{48}$$

where $\hat{R}_0$ is a lower bound on the cutoff rate. Since the result on the optimal input structure proved in Section III holds for all $\rho \in [0,1]$, we immediately conclude that the cutoff rate is achieved by an input whose amplitude has a discrete distribution while the phase is uniformly distributed. Indeed, numerical results indicate the optimality of a single-mass amplitude distribution in the low-SNR regime for high enough Rician factor (K) values. Motivated by this observation, we find closed-form expressions for both the cutoff rate lower bound and the Kuhn-Tucker condition when the input has single amplitude level.

*Proposition 1:* Fix $\rho = 1$. Then, the Kuhn-Tucker condition (24) for an input amplitude distribution with a single mass at $r = \sqrt{\alpha}$ becomes

$$\Phi_1(r) = \frac{2\sqrt{1+r^2}\sqrt{1+\alpha}}{2+\alpha+r^2} \exp\left(-\frac{\mathsf{K}r^2 + \mathsf{K}\alpha}{2(2+\alpha+r^2)}\right) I_0\left(\frac{\mathsf{K}r\sqrt{\alpha}}{2+\alpha+r^2}\right)$$
$$- \exp\left(-\frac{\mathsf{K}\alpha}{2(1+\alpha)}\right) I_0\left(\frac{\mathsf{K}\alpha}{2(1+\alpha)}\right) + \frac{\lambda}{2}(r^2 - \alpha) \geq 0 \quad (49)$$

for all $r \geq 0$ with equality at $r = \sqrt{\alpha}$. In the above formulation, $\lambda \geq 0$ is the Lagrange multiplier and for the special case in consideration, $\lambda$ is given by

$$\lambda = \frac{\mathsf{K}}{2(1+\alpha)^2} e^{-\frac{\mathsf{K}\alpha}{2(1+\alpha)}} \left(I_0\left(\frac{\mathsf{K}\alpha}{2(1+\alpha)}\right) - I_1\left(\frac{\mathsf{K}\alpha}{2(1+\alpha)}\right)\right) \quad (50)$$

where $I_1(\cdot)$ is the first-order modified Bessel function of the first kind.

*Proof:* Note that when $\rho = 1$, the LHS of the Kuhn-Tucker condition (24) for a discrete amplitude distribution with a single mass at $r = \sqrt{\alpha}$ is simplified to

$$\Phi_1(r) = \int_0^\infty \hat{g}(R, \sqrt{\alpha}) \hat{g}(R, r) \, dR - \int_0^\infty \hat{g}^2(R, \sqrt{\alpha}) \, dR + \frac{\lambda}{2}(r^2 - \alpha). \quad (51)$$

At this point, we note the integration result [12]

$$\int_0^\infty e^{-ax} I_0(b_1\sqrt{x}) I_0(b_2\sqrt{x}) \, dx = \frac{1}{a} e^{\frac{b_1^2 + b_2^2}{4a}} I_0\left(\frac{b_1 b_2}{2a}\right). \quad (52)$$

Using (52), we can easily show that $\int_0^\infty \hat{g}(R, \sqrt{\alpha}) \hat{g}(R, r) \, dR$ equals the first term in (49) and

$$\int_0^\infty \hat{g}^2(R, \sqrt{\alpha}) \, dR = \exp\left(-\frac{\mathsf{K}\alpha}{2(1+\alpha)}\right) I_0\left(\frac{\mathsf{K}\alpha}{2(1+\alpha)}\right), \quad (53)$$

proving the result in (49). Note that for optimality, $\Phi_1(\sqrt{\alpha}) = 0$ and $\Phi(r) > 0$ for all $r \in [0, \infty) - \{\alpha\}$. Therefore,

$$\left.\frac{d\Phi(r)}{dr}\right|_{r=\sqrt{\alpha}} = 0. \quad (54)$$

Solving the above equation provides the closed-form expression (50) of $\lambda$. □

Figures 1 and 2 plot $\Phi_1(\cdot)$ (49) as a function of $r$ for the single-mass discrete distribution function $F(r) = u(r - 0.3)$ in the Rician channel with $\mathsf{K} = 1$ and the distribution function $F(r) = u(r - 0.6)$ in the Rician channel with $\mathsf{K} = 2$, respectively. In both cases, we observe that the optimality conditions are satisfied and indeed the chosen single-mass distributions, together with uniform phase, are achieving the maximum in (48) in the respective channels. The following result provides a closed-form expression for $\hat{R}_0$ when a single-mass input is optimal.

*Corollary 1:* Assume that the input with uniform phase and discrete amplitude distribution with single-mass at $r = \sqrt{\alpha}$ is optimal, then

$$\hat{R}_0 = -\log \int_0^\infty \hat{g}^2(R, \sqrt{\alpha}) \, dR \quad (55)$$
$$= \frac{\mathsf{K}\alpha}{2(1+\alpha)} - \log I_0\left(\frac{\mathsf{K}\alpha}{2(1+\alpha)}\right) \quad (56)$$

The result follows from Proposition 1 where the integral in (55) has already been computed. Figure 3 plots $\hat{R}_0$ as a function of $\alpha$ in the Rician channel with $\mathsf{K} = 2$.

We have noted that the optimality of a single-mass amplitude distribution is observed for high enough Rician factors $\mathsf{K}$. The next result provides a precise value of $\mathsf{K}$ below which the single-mass input cannot be optimal for sufficiently small SNR values.

*Proposition 2:* If the Rician factor $0 \leq \mathsf{K} < -2 + 2\sqrt{2}$, then the single-mass amplitude distribution $F(r) = u(r - \sqrt{\alpha})$ cannot be optimal in the low-power regime.

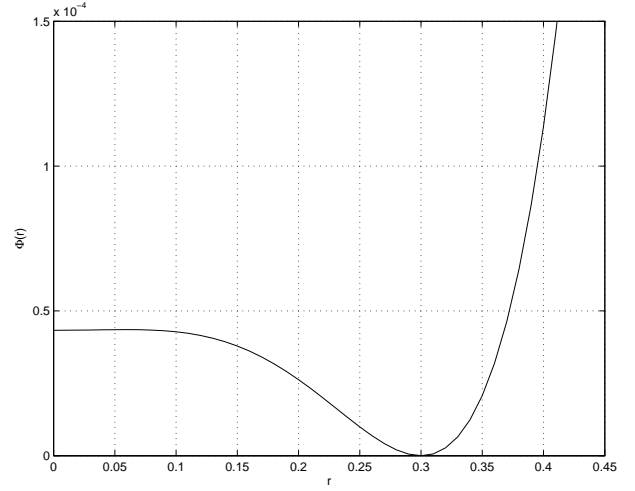

Fig. 1. The Kuhn-Tucker condition for $\mathsf{K} = 1$, $\alpha = 0.09$. $F(r) = u(r - 0.3)$. $\lambda = 0.3956$

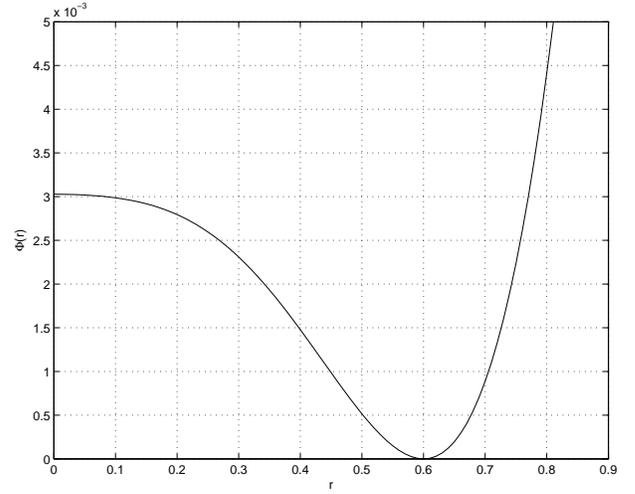

Fig. 2. The Kuhn-Tucker condition for $\mathsf{K} = 2$, $\alpha = 0.36$. $F(r) = u(r - 0.6)$. $\lambda = 0.3668$

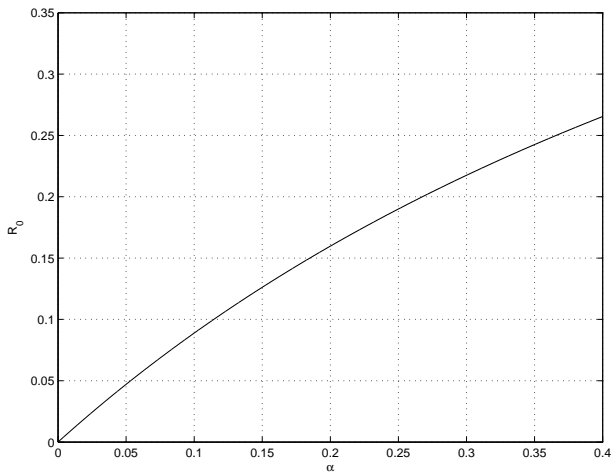

Fig. 3. $\hat{R}_0$ in nats vs. $\alpha$ in the Rician channel with $\mathsf{K} = 2$

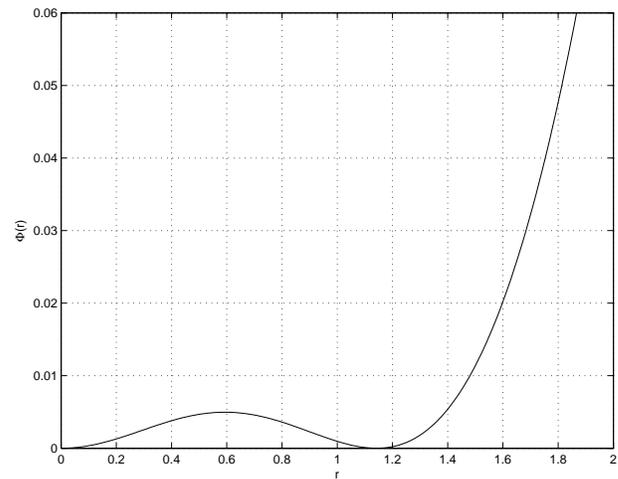

Fig. 4. The Kuhn-Tucker condition vs. $r$ for the Rician channel with $\mathsf{K} = 2$, $\alpha = 1$. The optimal input is $F(r) = 0.234\,u(r) + 0.766\,u(r - 1.142)$. $\lambda = 0.1422$. $\hat{R}_0 = 0.637$ nats

*Proof*: Consider the LHS of the Kuhn-Tucker condition (49) $\Phi_1(r, \alpha)$. Note that the dependence of $\Phi_1$ on $\alpha$ is explicitly indicated here. The following properties of $\Phi_1$ are easily verified:

$$\Phi_1(0, 0) = 0 \tag{57}$$

$$\left.\frac{\partial \Phi_1(0, \alpha)}{\partial \alpha}\right|_{\alpha=0} = 0 \tag{58}$$

$$\left.\frac{\partial^2 \Phi_1(0, \alpha)}{\partial \alpha^2}\right|_{\alpha=0} = \frac{\mathsf{K}^2 + 4\mathsf{K} - 4}{8}. \tag{59}$$

From the above properties, we see that if $\mathsf{K}^2 + 4\mathsf{K} - 4 < 0$ or equivalently $0 \leq \mathsf{K} < -2 + 2\sqrt{2}$, then $\Phi_1(0, \alpha) < 0$ for sufficiently small values of $\alpha$. Since the optimality condition is $\Phi_1(r, \alpha) \geq 0$ for all $r \geq 0$, the single-mass distribution cannot be optimal in this case. $\square$

We conjecture that the optimal input amplitude has a single mass for sufficiently low SNR values if $\mathsf{K} \geq -2 + 2\sqrt{2} = 0.8284$

We note that if the input is subject to both average and peak power constraints, we again observe that a single-mass input amplitude distribution is optimal in the low-SNR regime. Hence, the above analysis with the similar expressions follows for the average and peak power limited case. It is also interesting to note that in contrast to the requirement of flash signaling to achieve the capacity $C$ in unknown channels [4], $\hat{R}_0 < C$ in the low-power regime is achieved by an input with unit peak-to-average power ratio.

If the SNR is further increased, the numerical analysis indicates that $\hat{R}_0$ is achieved by a two-mass-point input distribution with one-mass at the origin. Figure 4 plots the Kuhn-Tucker condition for the two-mass-point input $F(r) = 0.234\,u(r) + 0.766\,u(r-1.142)$ at $\alpha = 1$ for the Rician channel ($\mathsf{K} = 2$). The optimality of this distribution is clearly seen in the figure.

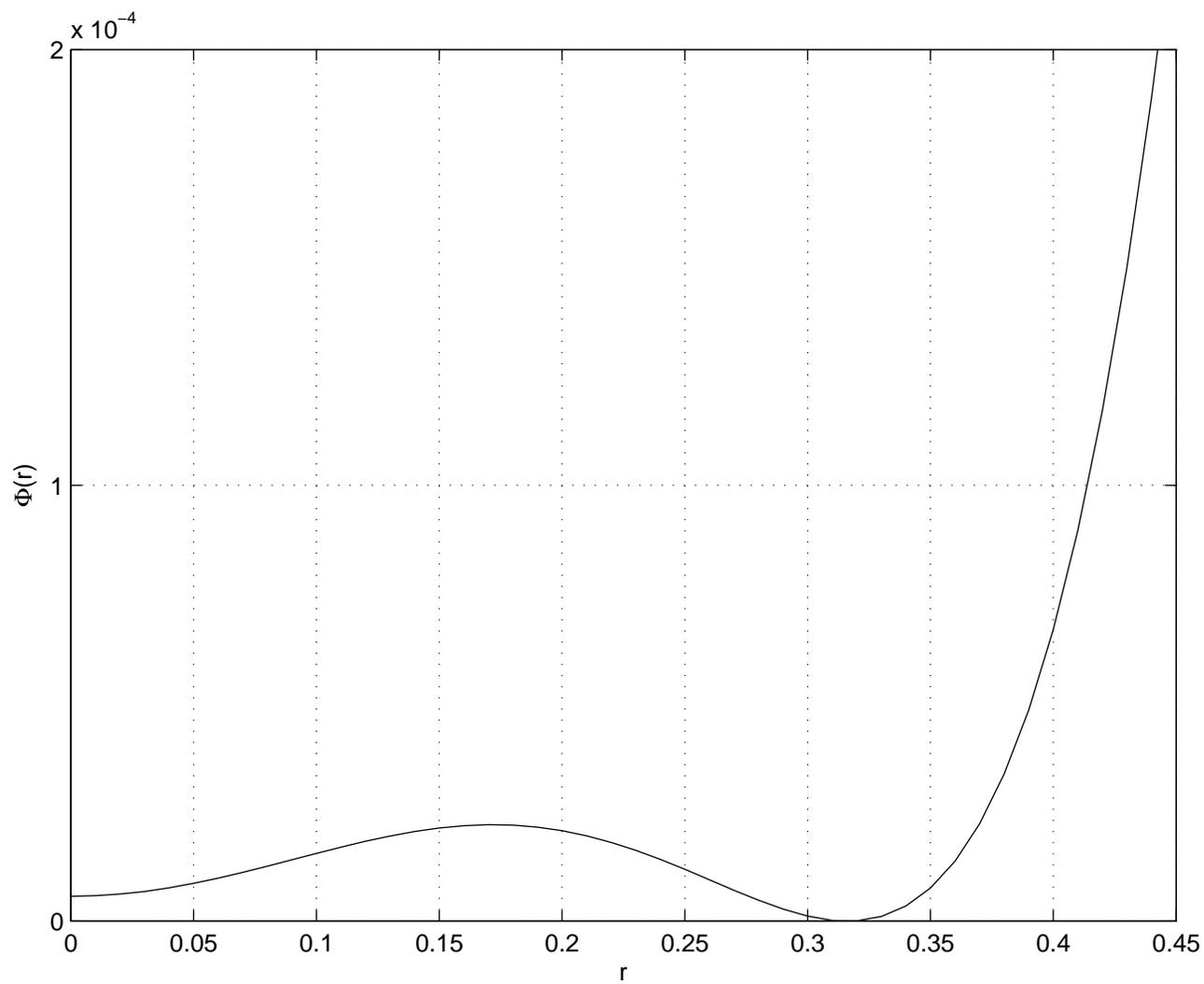